\documentclass[twocolumn,showpacs,preprintnumbers,amsmath,amssymb,floatfix]{revtex4}
\usepackage{graphicx}% Include figure files
\usepackage{dcolumn}% Align table columns on decimal point
\usepackage{bm}% bold math

\begin{document}

\newcommand{\etal}{\MakeLowercase{\textit{et al.}}} % "et al."

\title{Cosmic Ray Energy Spectrum from Measurements of Air Showers}

\author{
 T. K. Gaisser$^{1}$, T. Stanev$^{1}$ \& S. Tilav$^{1}$
}
\affiliation{
$^{1}$Bartol Research Institute and Department of Physics and Astronomy, University of Delaware, Newark, DE 19716, USA.
}

\begin{abstract}
This review focuses on high-energy cosmic rays in the PeV energy range and 
above.  Of particular interest is the knee of the spectrum around 3 PeV
and the transition from cosmic rays of Galactic origin to particles from 
extra-galactic sources.  Our goal is to establish
a baseline spectrum from $10^{14}$ to $10^{20}$~eV by combining
the results of many measurements at different energies.  In combination with 
measurements of the nuclear composition of the primaries, the shape of
the energy spectrum places constraints on the number and spectra of
sources that may contribute to the observed spectrum.

\end{abstract}

\pacs{...}

\maketitle

%Begin the section.
\section{Introduction}
\label{sec:intro}

In the 100 years since Victor Hess's balloon flight that
marks the discovery of cosmic rays, a great deal has been 
learned about their composition, propagation and sources.
We know that most particles originate from sources in the 
local galaxy, having spent on average $10^7$ years in diffusive motion
in the interstellar medium (ISM) and the galactic halo before
being lost to intergalactic space~\cite{lifetime}.  
We know that the power requirement
to maintain the intensity of cosmic rays at a near constant value
is a few per cent of the energy available in supernova explosions
and that there is a well developed theory of non-linear, diffusive
acceleration by shocks driven by expanding supernova remnants 
(SNR)~\cite{MalkovDrury}.
We know that diffusion in the interstellar medium depends on energy in such a way that
higher energy nuclei escape more quickly than lower energy ones, but
the shape of this energy dependence is still uncertain at high energy~\cite{Obermeier}.

Measurements of diffuse gamma radiation from $<1$ to $>100$ GeV
by the Fermi Satellite~\cite{Fermi} confirm this general picture of the origin
and propagation of galactic cosmic rays.  In the TeV range, ground
based gamma-ray telescopes have observed emission of $>$TeV gamma-rays
from supernova remnants, as expected if there is particle acceleration
to energies of $>10$~TeV~\cite{Holder}.  However, because electrons are more
efficient radiators than protons, the relation of these observations
to the fluxes of high-energy protons and nuclei observed at Earth
is not completely clear.  In particular, the maximum energy to which
protons and nuclei can be accelerated by supernova remnants is not yet directly established.

\begin{figure*}[!th]
%  \vspace{5mm}
  \centering
\includegraphics[width=15cm]{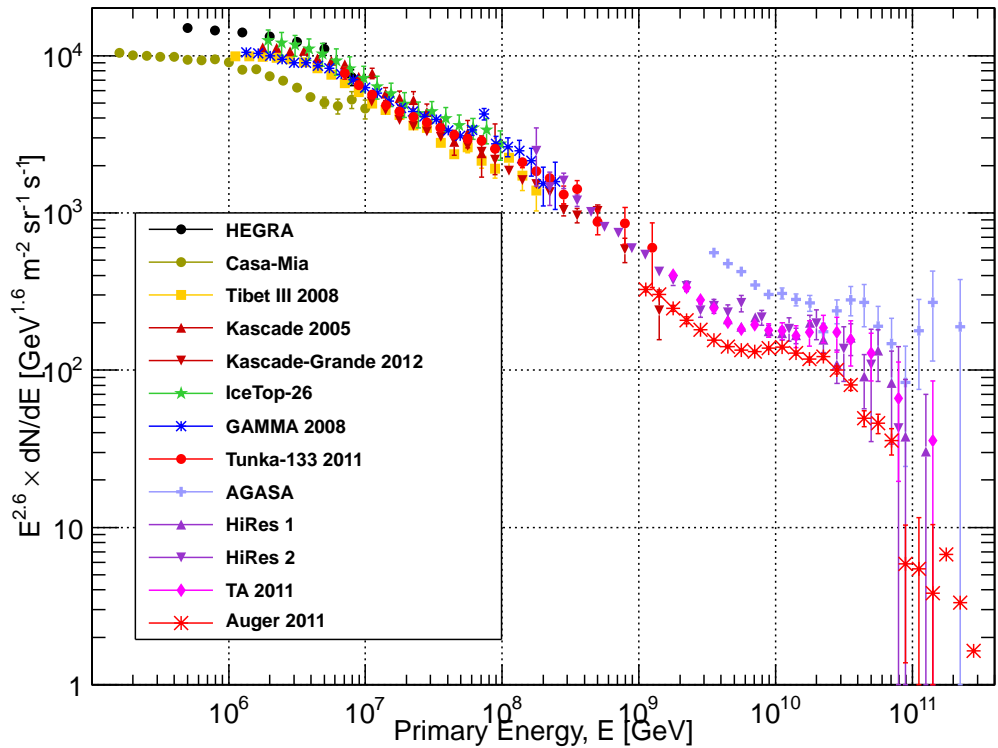}
  \caption{All particle cosmic ray spectrum from air shower experiments.
  (References in text.)}
  \label{allpart}
 \end{figure*}

The picture outlined above is driven by direct measurements
of cosmic rays with detectors near the top or outside of the atmosphere.
Measurements with spectrometers in high-altitude balloons
and satellites measure nuclear and isotopic composition 
in the GeV energy range and spectra of individual elements 
up to TeV/nucleon~\cite{review1}.  Instruments with calorimeters and 
transition radiation detectors extend the energy measurements
to $\sim 100$ TeV per nucleus.  Recent measurements
by CREAM~\cite{CREAM,discrepant} and PAMELA~\cite{PAMELA} support
earlier observations by JACEE~\cite{JACEEpHe,JACEE} and ATIC~\cite{ATIC}
that the spectrum of helium is somewhat harder than that of protons.
The PAMELA measurements show that the spectrum of both protons and
helium become harder above a rigidity of 200 GV.  The CREAM data suggest
that this is also the case for heavier nuclei~\cite{discrepant}.
Further new data are expected soon from the AMS-2 detector that by now has
 worked on the International Space Station for more than one year.  

The cosmic-ray flux above 100 TeV amounts to about 5 particles per square meter
per steradian per day.  Therefore, direct measurement of the cosmic-ray spectrum
above this energy with balloon borne or satellite detectors is difficult.  
The highest direct measurement so far is still the series of measurements
by Grigorov \etal~carried on \textit{Proton} satellites~\cite{Grigorov}, 
which extends to $\sim$PeV. 
Data for the spectrum above a few hundred TeV comes from large air shower
arrays on the ground.  The air-shower experiments observe the cascades of
secondary particles in the atmosphere initiated by the interaction of the
high energy primary particles.  In such indirect measurements, the
information about composition is limited (at best) to determining
the relative abundances of the main groups of nuclei.  Moreover, the
relation between the observed signal at the ground and the primary energy
depends on the mass of the primary nucleus.  In addition, the relation between
the observed signal and the primary energy depends on the model(s) of hadronic
interactions used to interpret the cascades.  Given the large uncertainties
in energy assignment, it is not surprising that different measurements
give different results in the same or overlapping energy regions.  
Another complexity is that techniques for measuring the shower at
the ground differ from one experiment to the next (for example, scintillators
compared to water Cherenkov tanks).  In this
review we therefore use observed features in the energy spectra 
to cross-calibrate results from different arrays.

Since air shower measurements are calorimetric in nature, the natural energy
variable to use is total energy per nucleus.  This contrasts with the direct
measurements, which generally use energy per nucleon.  
The latter is the natural variable for studying propagation of nuclei because energy per nucleon is
conserved in spallation.  A variable different from both of these is relevant
for interpreting observed spectra.  This is magnetic rigidity, defined as
\begin{equation}
R\,=\,\frac{Pc}{Ze},
\label{rigidity}
\end{equation}
where $P$ is the total momentum of a nucleus
and $Ze$ its electrical charge.
Particles with the same rigidity and injection vector follow identical trajectories in
a given magnetic field configuration.  Rigidity is therefore the
appropriate variable for interpreting changes in spectrum due to propagation
and acceleration in magnetic fields.  In particular, as first pointed out
by Peters~\cite{Peters}, 
if there is a maximum energy to which protons can be accelerated
in a source, then the protons will cutoff first, followed by helium, carbon, $\ldots$ according to 
\begin{equation}
E_{\rm max}(Z)\,=\,Ze\times R_c\;=\;Z\times E_{\rm max}(Z=1).
\label{Peters}
\end{equation}

The outline of the paper is as follows.  We first describe briefly the different types of
air shower experiments and summarize the data from each of the selected measurements.  
We then use features in the energy spectrum observed by different experiments to construct 
a tentative all-particle energy spectrum from 
$10^{14}$ to $10^{20}$~eV.  The most prominent features are the knee around $3\times 10^{15}$~eV 
and the ankle around $10^{19}$~eV, both prominent in Fig.~\ref{allpart}.  As
justification for this approach, we note that, to a high degree of accuracy,
there must be a single spectrum at Earth.  In the third major section of the
paper we use this constructed all-particle spectrum as a template for discussing measurements of composition and possible implications for different sources.
In this section we will assume the validity of the Peters cycle as written
in Eq.~\ref{Peters} as a constraint on the energy dependence of different
nuclear components.  We describe two fits to the data, each of which has three populations
of particles with contrasting assumptions about the rigidity cutoff for each population.
The first two populations represent cosmic rays from galactic sources and the third
population is an extragalactic component.  Each population contains several groups of
nuclei with assumed spectral indices as adjustable parameters.

\begin{table*}
 \begin{tabular}{|l|c|r|r|} \hline
 {Array} & {Overburden (g/cm$^2$} & {type} & {Energy range} \\ \hline \hline
 {Tibet~\cite{Tibet2}}   & 559    & Scintillator & 1 - 200 PeV  \\ \hline
 {Akeno~\cite{Akeno}} & 909   & Scintillator/muon & 0.5 PeV - 5 EeV  \\ \hline
 {AGASA~\cite{AGASA}} & 909   & Scintillator & 4-200 EeV  \\ \hline
 {HEGRA~\cite{HEGRA}} & 755  & Air Cherenkov & 0.5-10 PeV \\ \hline
 {TUNKA~\cite{TUNKA}} & 938  & Air Cherenkov & 7-1000 PeV  \\ \hline
 {Kascade~\cite{Kascade}} & 1022 & Scintillator/muon & 2-90 PeV  \\ \hline
 {Kascade-Grande~\cite{KGrande2}} & 1022 & Scintillator/muon & 1-1400 PeV  \\ \hline
 {GAMMA~\cite{GAMMA}}& 700 & Scintillator/muon & 3-200 PeV \\ \hline
 {IceTop~\cite{IceTop}} & 680 & Ice Cherenkov & 1-1000 PeV  \\ \hline
 {IceCube~\cite{IceCube}} & 680 & Ice Cherenkov (surface) + deep muons & 1-1000 PeV  \\ \hline
 {CASA-MIA~\cite{CASA-MIA}} & 860    & Scintillator + muon counters & 0.1-10 PeV  \\ \hline
 {Fly's Eye~\cite{FlyEye}} & 860 & Air Fluorescence & 1-100 EeV  \\ \hline
 {Hi-Res~\cite{HiRes}} & 845  & Air Fluorescence & 0.2-100 EeV  \\ \hline
 {Telescope Array~\cite{TA}} & 845 & Hybrid & 2-140 EeV  \\ \hline
 {Auger~\cite{Auger}} & 845 & Hybrid & 1-280 EeV  \\ \hline
 \end{tabular}
 \label{tab1}
 \caption{{List of selected air shower detectors.  
 \textit{Hybrid} refers to air fluorescence telescopes looking over a surface array.}}
\end{table*}

\section{Air shower experiments}

Air shower detectors fall into several categories depending on the type
of sensors used and on the altitude of the array.  Scintillators such as those
used in the Akeno array detect charge particles, which are mostly electrons
and positrons with a fraction of order 10\% of muons.  In some arrays, 
the muon component can be distinguished from the electromagnetic component, either by a
second layer of scintillator with an absorber between the top and bottom layers
 (as at Akeno~\cite{Akeno}) or by separate, shielded, muon detectors (as
in CASA-MIA~\cite{CASA-MIA} and Kascade~\cite{Kascade}). 
The surface array at Auger~\cite{Auger} and IceTop~\cite{IceTop} both use 
tanks filled with water or ice to detect the Cherenkov radiation produced
by charged particles in the tanks.  The method, pioneered in the Haverah Park array~\cite{Haverah}, 
is sensitive to photons in the shower that convert in
the tanks as well as to the less numerous $e^\pm$ and muons.  

 HEGRA~\cite{HEGRA} and TUNKA~\cite{TUNKA} 
both use unshielded photomultipliers looking up at the night sky to detect
the atmospheric Cherenkov light generated by the atmospheric cascade.
In these arrays the depth of maximum,
and hence the chemical composition of the cosmic rays, is judged by the lateral
distribution of the Cherenkov light at the ground.  The atmospheric
fluorescence technique, first successfully used by Fly's Eye~\cite{FlyEye},
traces the isotropically emitted nitrogen fluorescence lines excited by
the passage of charged particles through the atmosphere.  This method,
after correcting for propagation through the atmosphere, maps the longitudinal
development of each shower in the atmosphere.  It therefore comes closest to
providing a direct, calorimetric measurement of the energy of each event.
By tracing each shower profile, a depth of maximum is assigned to each event.
The distribution of depths of maximum is sensitive to primary composition.
The atmospheric Cherenkov detectors provide depth of maximum and composition
information in the region of the knee up to about 100 PeV, while the Fly's
Eye type detectors provide this information around 1 EeV and above.

A partial list of air shower detectors and information about them is given
in Table~\ref{tab1}.  An account of the history of the air shower method
and the development of the various techniques is given in the review
of Kampert \& Watson~\cite{KW}.  The current status of spectrum and composition
is reviewed in the paper of Kampert \& Unger~\cite{KU}.

 \begin{figure*}[!th]
\includegraphics[width=\columnwidth]{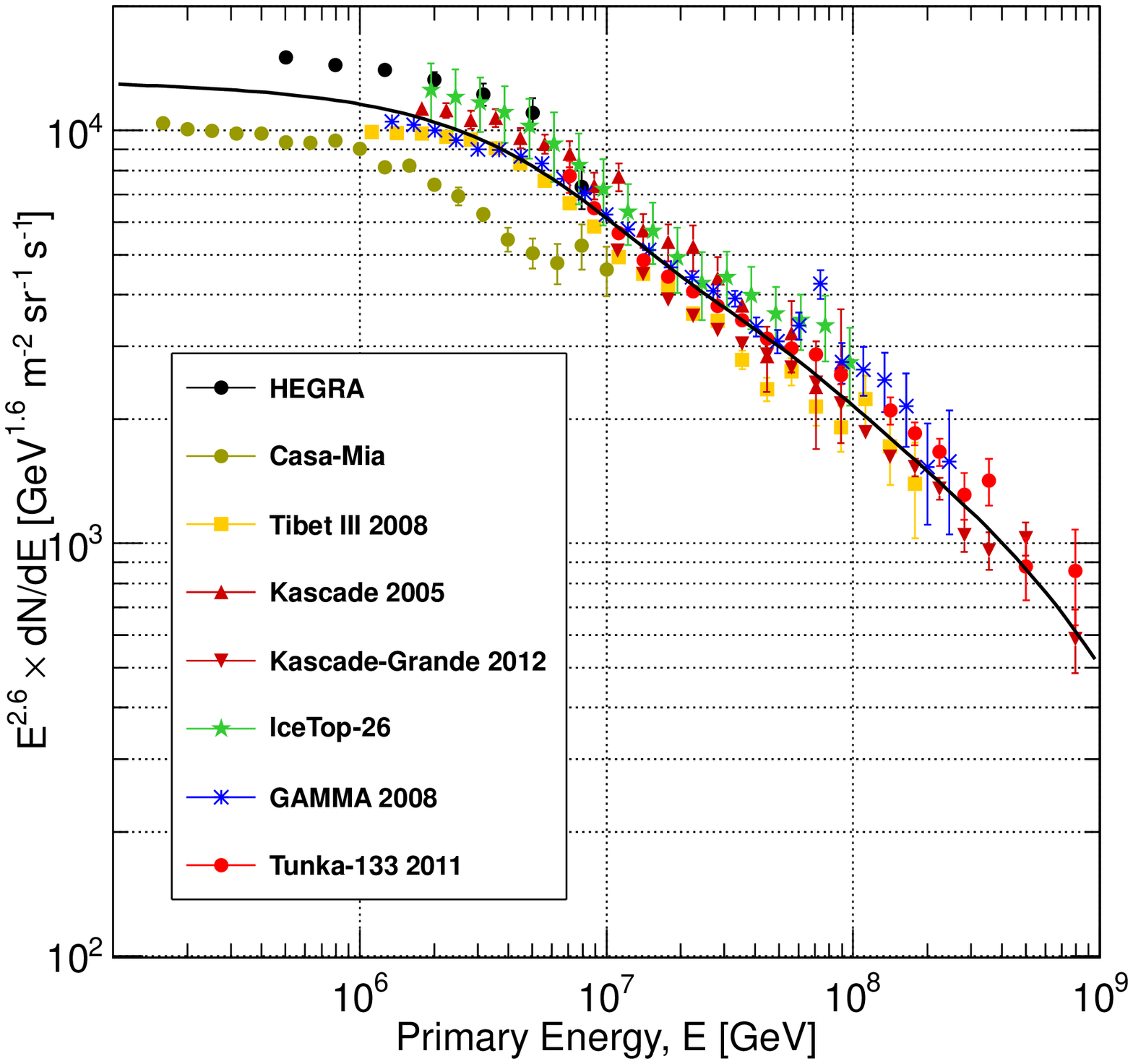}\,\includegraphics[width=\columnwidth]{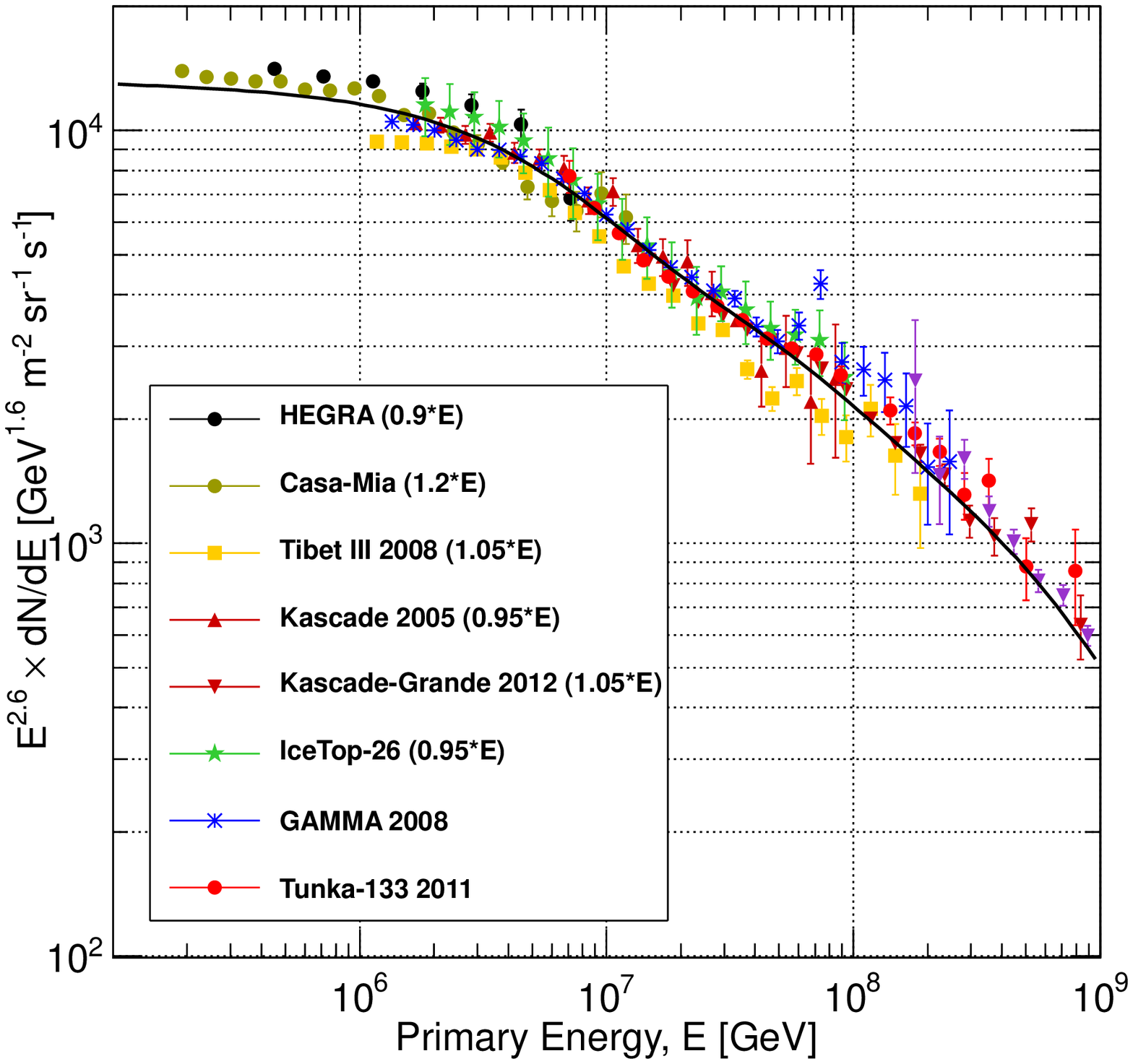}
  \caption{Data from kilometer-scale air shower experiments
  for the spectrum up to one EeV, including the knee of the spectrum.
  Left: data as presented; Right: data replotted with energies shifted
  as shown in the labels.}
  \label{knee}
 \end{figure*}

\section{Finding a single all-particle spectrum}

The technique of using a feature of the energy spectrum to inter-calibrate
different measurements was emphasized by Berezinsky, Gazizov \& Grigorieva
in their model of the ankle~\cite{BGG}.  They explain the ankle (which appears
as a ``dip" when the spectrum is plotted as E$^3$dN/dE) as being the result
of energy loss by protons from sources at cosmological distances to
 $e^\pm$ pair production in the cosmic microwave background (CMB). 
 They compared measurements
 of Yakutsk~\cite{Yakutsk}, a preliminary version of HiRes data~\cite{HiRes}, and a combination of
 Akeno (high energy)~\cite{Akeno2} and AGASA~\cite{AGASA}, shifting the
 energy scales respectively by $0.73$, $1.2$ and $0.9$.  In a recent
 review~\cite{ABG}, this approach is refined, extended to include more 
 recent data~\cite{Auger2},\cite{TA}, and compared to three different models of the
 transition from Galactic to extragalactic cosmic rays.
 At lower energy, the position of the knee has been used to provide relative
 calibrations of different measurements in the energy range from
 $3\times 10^{14}$ to $\sim 10^{16}$~eV~\cite{EW}.  
 
\begin{figure*}[!th]
\includegraphics[width=\columnwidth]{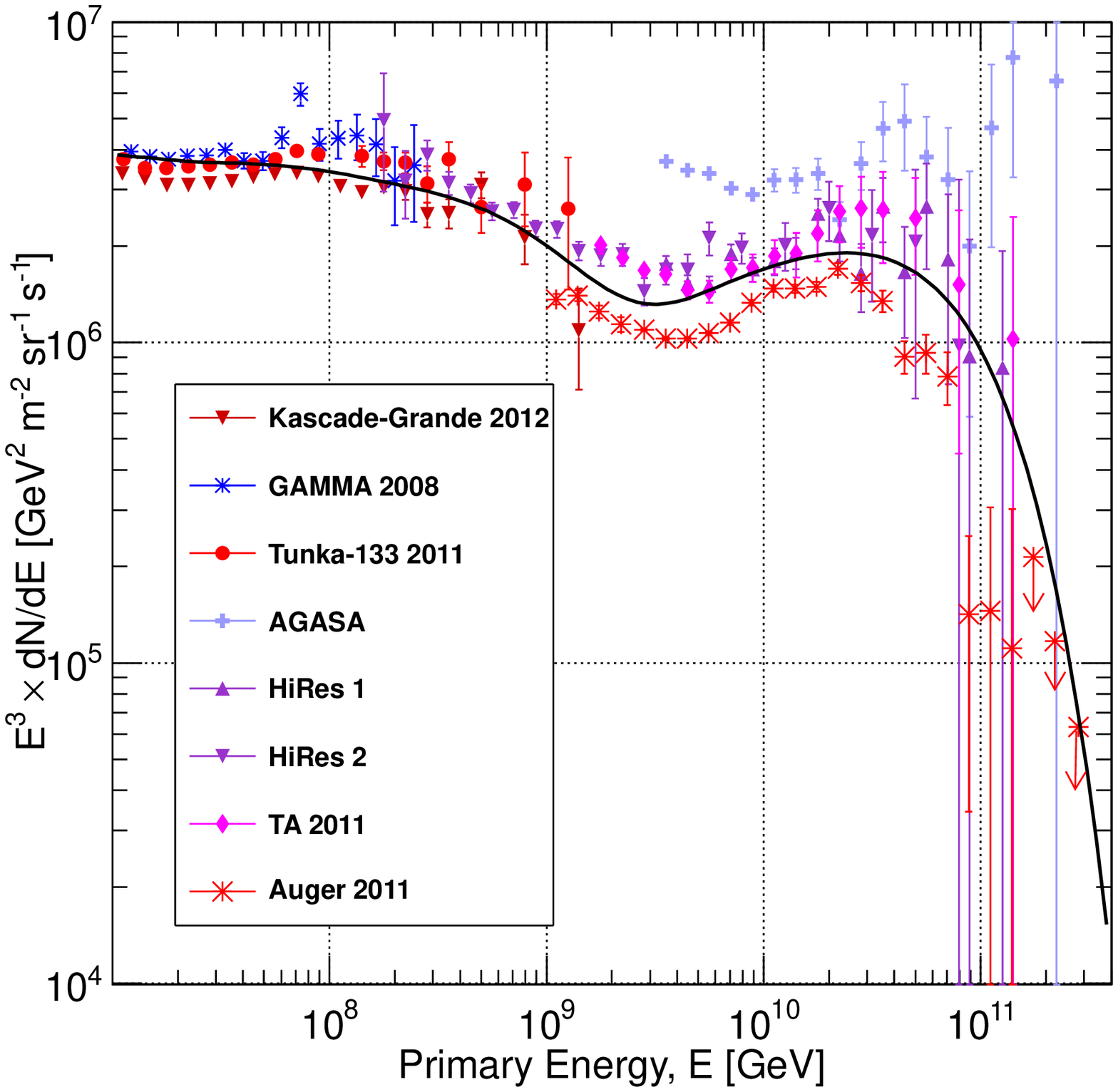}\includegraphics[width=\columnwidth]{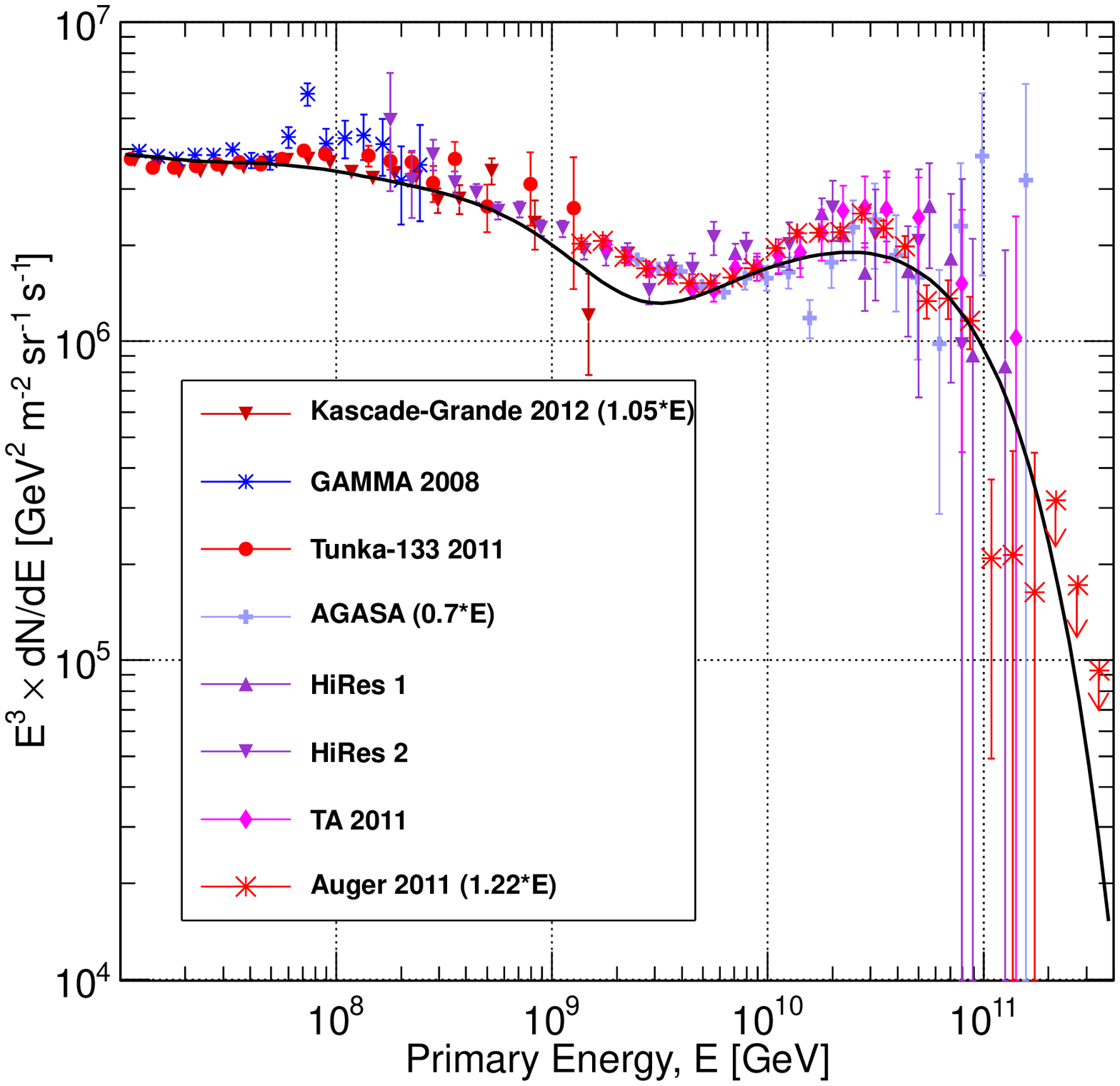}
  \caption{Data from giant air shower detectors.  Left: 
  Data from AGASA~\cite{AGASA}, HiRes~\cite{HiRes}, Auger~\cite{Auger}
  and Telescope Array~\cite{TA};  Right: same with data of AGASA shifted
  down in energy by 0.7 and Auger shifted up in energy by 1.22.}
  \label{ankle}
 \end{figure*}

 In this section,
 we show air shower data in two overlapping energy regions, $10^{14}$ to $10^{18}$~eV
 and $10^{16}$ to $10^{21}$~eV.  The first includes the knee and
 the second the ankle and the end of the spectrum.  In these plots
 the lines show the all particle spectrum proposed in 
 Ref.~\cite{GaisserAPP}, which we discuss in the following section.
It is important to keep in mind the amplification that occurs when data are
plotted as $E^{\gamma+1}{\rm d}N/{\rm d}E$, as is commonly done to display
relatively small deviations from the underlying power law structure of
the cosmic-ray spectrum.  Air shower experiments typically measure and present
there results as number of events per logarithmic bin of energy.  
Uncertainties in reconstructed energy generally scale with energy, so the
resolution is expressed as $\delta E/E\,=\,\delta\ln(E)$.  If the energy scale
is shifted by an amount $\delta\ln(E)$, then each point moves on a log-log plot
by a distance $\sqrt{1+\gamma^2}\,\delta\ln(E)$ at an angle elevated from
the horizontal by $\theta\,=\,\tan^{-1}(\gamma)$.  For $\gamma=1.6$ as in Fig.~\ref{knee} 
the shift is amplified by $1.89$ and for $\gamma=2$ as in
Fig.~\ref{ankle} the factor is $2.24$.  Thus, differences among data sets are not 
as large as they appear when multiplied by a power of the energy.

\subsection{The knee region}

To study the knee region we select several measurements that have similar 
(though not identical) structure to each other, but are offset to some extent.
The left panel of Fig.~\ref{knee} shows measurements of CASA-MIA~\cite{CASA-MIA}, 
KASCADE~\cite{Kascade}, HEGRA~\cite{HEGRA} and
Tibet~\cite{Tibet2}, all of which show a bending corresponding to
the knee of the spectrum.  Three higher energy measurements, which start above the
knee and extend to $10^9$~GeV
(Tunka~\cite{TUNKA}, GAMMA~\cite{GAMMA}
and KASCADE-Grande~\cite{KGrande2}) are also included.
Tibet, being a closely spaced array at high altitude has data down to
$100$~TeV.  CASA-MIA was also a closely spaced array at intermediate altitude.

The right panel of Fig.~\ref{knee} shows the spectra replotted with energies
shifted by factors as noted in the labels on the plot.  Some differences in
shape remain (particularly between CASA-MIA and Tibet in the knee region),
but it is possible with moderate shifts to bring the data into substantially
better agreement.  Since the location of the knee is not fixed a priori, it is
not possible to decide which experiment has the best absolute calibration.

\subsection{Ankle region}

Data up to the highest energies are collected in Fig.~\ref{ankle}.  Data in
the region of the ankle are from the giant air shower detectors AGASA~\cite{AGASA}, 
HiRes 1 and 2~\cite{HiRes}, Auger~\cite{Auger2} and Telescope Array~\cite{TA}.
The data below $10^9$~GeV from Tunka, GAMMA and KASCADE-Grande are also included.
The line shown is the model of Reference~\cite{GaisserAPP} with an extragalactic 
proton component.  Larger shifts are needed to bring the different measurements
into better agreement than in the case of the knee region.
In their recent review~\cite{ABG},
the Auger data has been shifted up by a factor 1.22, which we also use.
This brings the data into agreement in the decade between $10^9$ and
$10^{10}$~GeV.

\section{Approaches to describing the spectrum}

Although it is possible to make a model in which the entire observed cosmic-ray spectrum 
comes from sources in our galaxy~\cite{Kusenko}, 
it is generally accepted that the
knee is in some way associated with the beginning of the end of a population
particles accelerated by sources in the Milky Way and that the highest energy
particles are from extragalactic sources.  
In this section we explore two rather different realizations of this
basic idea.  In both cases we follow the reasoning of
Peters~\cite{Peters} by assuming
that the knee and other features of the
primary spectrum depend on magnetic rigidity as defined in Eq.~\ref{rigidity}.
 The motivation for this assumption is that both acceleration
and propagation in models that involve collisionless diffusion
in magnetized plasmas depend only on rigidity.  
The first evidence for a Peters cycle associated with the knee of
the cosmic-ray spectrum comes from the unfolding analysis of
measurements of the ratio of low-energy muons to electrons at the
sea level with the KASCADE detector~\cite{Kascade}.  They found
that the knee occurred earlier for protons and helium and later 
for heavier nuclei.  The same Peters cycle pattern seems to
occur also in the hardening of spectrum observed recently
around 200~GV as reported in Refs.\cite{discrepant} and~~\cite{PAMELA}.

\subsection{Hillas model}

The model of Ref.~\cite{GaisserAPP}
is an attempt to implement the model of Hillas~\cite{Hillas} in which the knee
represents the end of the spectrum of cosmic rays accelerated by supernova
remnants in the Milky Way and the ankle represents the transition to particles
from extra-galactic sources.  This picture depends on the
amplification of magnetic fields by the turbulence associated with
non-linear diffusive shock acceleration~\cite{Bell}.  Support for the
presence of magnetic field amplification by a factor of 100 above
the level the interstellar medium comes from the narrow rims
of synchrotron radiation by electrons observed at the edges of some
SNR~\cite{narrow}.  With fields of order $100~\mu$Gauss, acceleration of
protons to energies $E_{\rm max}\sim 3 \times 10^6$~GeV is possible
given the size and expansion rate of SNR~\cite{Blasi}.  In
this situation it is natural
to associate the knee with the maximum energy for the bulk of the galactic
cosmic rays.

\begin{table}[thb]
\begin{center}
\begin{tabular}{l|ccccc}  \hline
&  p & He & CNO & Mg-Si & Fe \\ \hline  
 Pop. 1: & 7860 & 3550 & 2200 & 1430 & 2120  \\ 
$R_c=4$~PV &1.66 1 & 1.58  &   1.63   &   1.67     &   1.63  \\ \hline
Pop. 2:  & 20 & 20  &   13.4    &   13.4     &  13.4  \\
$R_c=30$~PV & 1.4 & 1.4 & 1.4 & 1.4 & 1.4  \\ \hline
Pop. 3: &1.7 &1.7  &   1.14    & 1.14   & 1.14 \\ 
 $R_c=2$~EV  & 1.4 & 1.4 & 1.4 &  1.4 & 1.4  \\ \hline \hline
Pop. 3(*): & 200 &  0.0 &   0.0   &  0.0  & 0.0      \\ 
 $R_c=60$~EV  & 1.6 &  &  &  &  \\ \hline \hline
 \end{tabular}
\caption{Cutoffs, normalization constants ($a_{i,j}$) and integral
spectral indexes ($\gamma_{i,j}$)
for Eq.~\ref{ModelH3} for the implementation of the Hillas model (H3a) in which
all populations are mixed.  
In the bottom part of the table population 3(*) consists of protons only (H4a).
}
\label{tab2}
\end{center}
\end{table}

If the ankle signals the transition to extragalactic cosmic rays,
and the cutoff for the SNR component occurs at a rigidity of several PV,
then there is a gap between the knee and the ankle that has to be filled
in by a higher energy galactic component, which Hillas calls ``component B.''
In this case there would be at least three populations of particles.
There could of course be many more components in a more realistic 
picture in which different classes of sources, or even individual
sources have different individual characteristics.  For this reason
a three population model is a minimal assumption in case the transition
to extra-galactic cosmic rays occurs at the ankle.

\begin{figure*}[thb]
\includegraphics[width=\columnwidth]{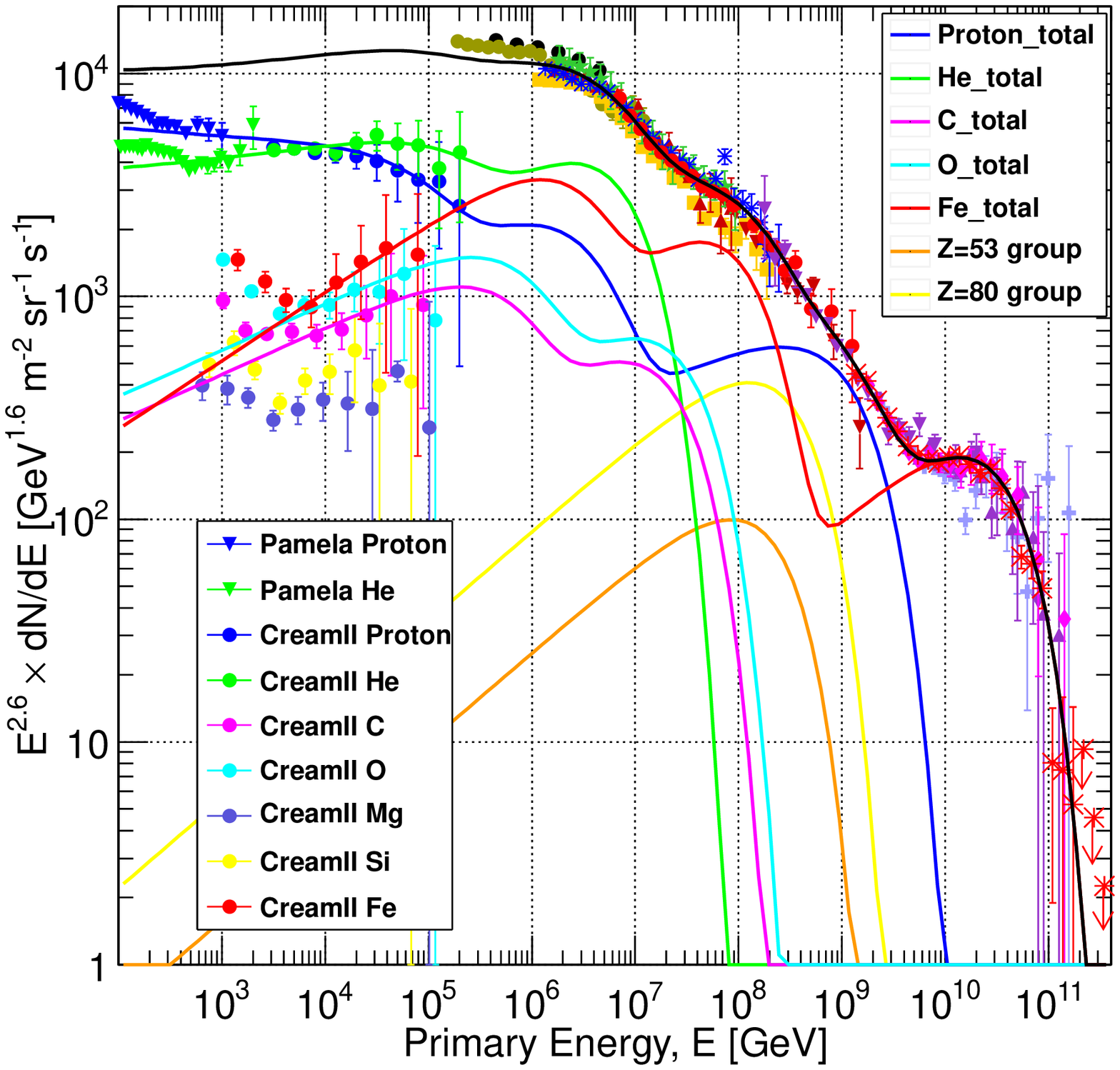}\includegraphics[width=\columnwidth]{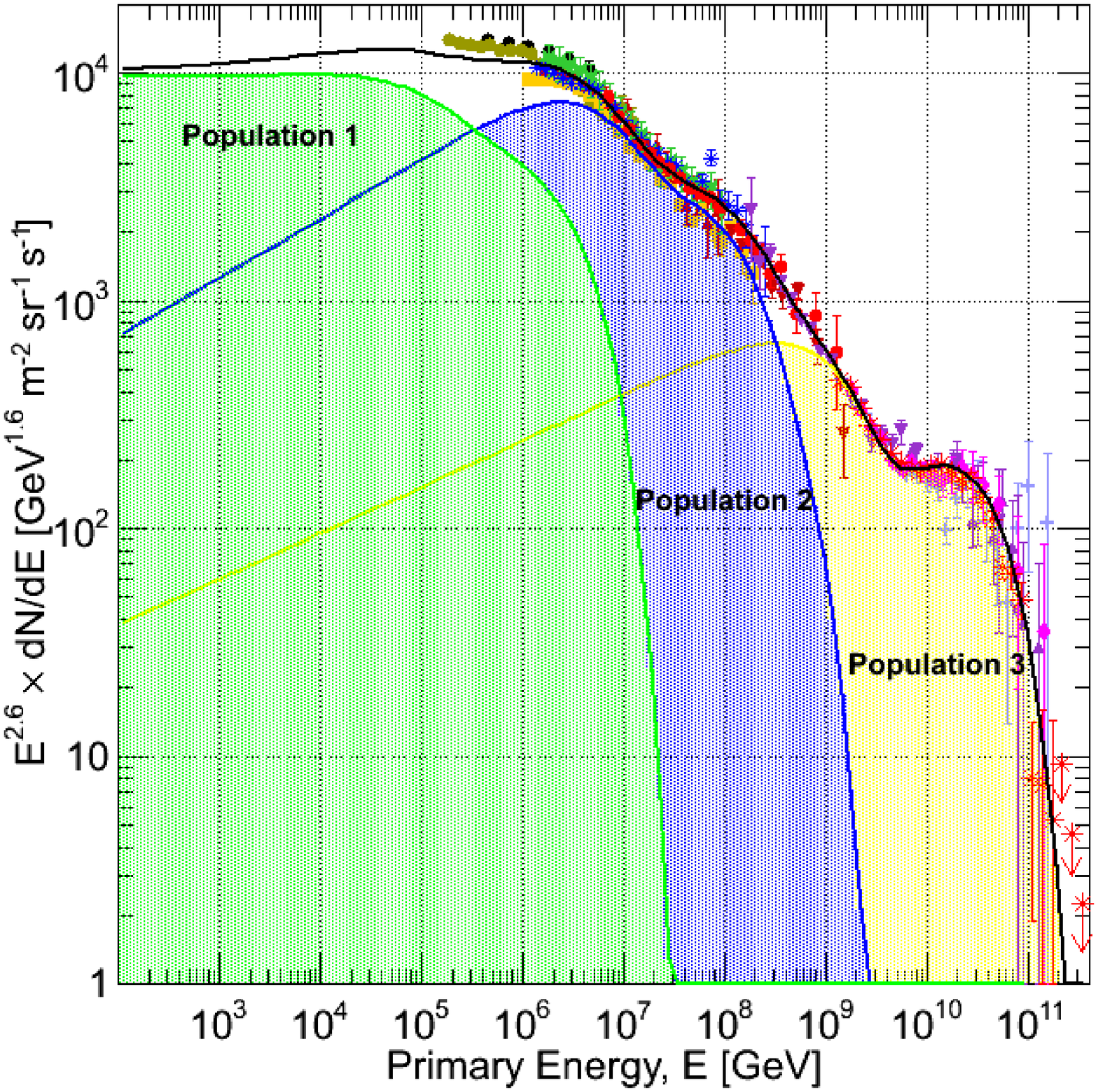}
  \caption{Overview of the spectrum from below the knee to the ankle with the fit of Table~\ref{tab3}.  
  Air shower data shifted 
  as in Figs.~\ref{knee} and~\ref{ankle}.  Left: lines showing individual groups of nuclei from all populations compared
  to data from PAMELA~\cite{PAMELA} and CREAM~\cite{CREAM} at low energy.  Right: shaded regions show
  the overlapping contributions of the three populations.
  }
  \label{global}
 \end{figure*}

This three population picture is implemented in the model of Ref.~\cite{GaisserAPP}
by assuming that each of the three components ($j$)
contains all five groups of nuclei and cuts off
exponentially at a characteristic rigidity $R_{c,j}$.  Thus the all-particle
spectrum is given by
\begin{equation}
\phi_i(E)\;=\;\Sigma_{j=1}^3\,a_{i,j}\,
E^{-\gamma_{i,j}}\times \exp\left[-{E\over Z_i R_{c,j}}\right].
\label{ModelH3}
\end{equation}
The spectral indices for each group and the normalizations are given explicitly in
Table~\ref{tab2}.  The parameters for Population 1 are from Refs.~\cite{CREAM,discrepant},
which we assume can be extrapolated to a rigidity of $4$~PV to describe the knee.
In Eq.~\ref{ModelH3} $\phi_i$ is d$N/$d$\ln E$ and $\gamma_i$ is the integral spectral index.
The subscript $i=1,5$ runs over the standard five groups (p, He, CNO, Mg-Si and Fe), and the 
all-particle spectrum is the sum of the five.
  This model is plotted
as the solid line in Figs.~\ref{knee} and~\ref{ankle}.

  \begin{table*}[thb]
\begin{center}
\begin{tabular}{l|ccccccc}  \hline
&  p & He & C & O & Fe & $50<Z<56$ & $78<Z<82$\\ \hline  
 Pop. 1: & 7000 & 3200 & 100 & 130 & 60 & & \\ 
$R_c=120$~TV &1.66 1 & 1.58  &   1.4   &   1.4     &   1.3 & &  \\ \hline
Pop. 2:  & 150 &65  &   6    &   7     &   2.3 & 0.1 & 0.4  \\
$R_c=4$~PV & 1.4 & 1.3 & 1.3 & 1.3 & 1.2 & 1.2 & 1.2 \\ \hline
Pop. 3: &14 &   &       &    & 0.025     &    & \\ 
 $R_c=1.3$~EV  & 1.4 &  &  &  & 1.2 &  & \\ \hline \hline
 Pop. 2*:  & 150 &65  &   6    &   7     &   2.1& 0.1 & 0.53  \\
$R_c=4$~PV & 1.4 & 1.3 & 1.3 & 1.3 & 1.2 & 1.2 & 1.2 \\ \hline
Pop. 3*: &12 &   &      &    & 0.011     &    & \\ 
 $R_c=1.5$~EV  & 1.4 &  &  &  & 1.2 &  & \\ \hline 
Pop. 4*: &1.2 &   &       &   &     &    & \\ 
 $R_c=40$~EV  & 1.4 &  &  &  &  &  & \\ \hline \hline
\end{tabular}
\caption{Global Fit results for the cutoffs, normalization constants ($a_{i,j}$) and integral
spectral indexes ($\gamma_{i,j}$)
for Eq.~\ref{ModelH3}.  In the bottom part of the table(*) populations 2 and 3 are slightly modified
to accommodate a Population 4 of protons to bring $<\ln(A)>$ down to the observed level in Fig.~\ref{lnA}.}
\label{tab3}
\end{center}
\end{table*}

 \subsection{An alternative picture and global fit}
 
 Spectra for the second fit are given by the same Eq.~\ref{ModelH3}
 but with qualitatively different parameters, as given in Table~\ref{tab3}.
 In particular, the first population has a much lower 
 cutoff of $R_c=120$~TV. 
 This description is related to the significantly harder spectra
 assumed for the first population.  Each component in the first population is fitted only
 above $R_c=200$~GV, after the spectra hardening noted in Refs.~\cite{discrepant}
 and~\cite{PAMELA}.  With these harder spectra (as compared to Table~\ref{tab2}),
 the heavy components cannot be extended past the knee region.  It is interesting to
 note that $R_c\approx 100$TV is the classical result for the expected maximum
 energy of supernova remnants expanding into the interstellar medium with an
 un-amplified magnetic field of a few $\mu$Gauss~\cite{LagageCesarsky}.

The spectrum with the parameters of Table~\ref{tab3} is shown in Fig.~\ref{global}
from below the knee to the ankle.  The contributions
of individual groups of nuclei are shown, as well as the spectra
of nuclei from CREAM~\cite{discrepant}.  We note that the
bump in the spectrum around $10^{17}$~eV corresponds with the
``iron knee''  reported by KASCADE-Grande in their electron
rich sample~\cite{KGrande} and also noted by GAMMA~\cite{GAMMA}.
A tendency for increasing mass above the knee has been
noted for a long time (for example by CASA-MIA~\cite{CASA-MIA2}), which
seems now to be confirmed with higher resolution.

Another noteworthy feature is the possibility illustrated in this
fit of explaining the ankle as a Peters cycle containing only protons
and iron.  This possibility is also suggested in Ref.~\cite{ABG} as an
example of their ``disappointing'' model~\cite{disappointing} of the end of the cosmic-ray
spectrum.  Such a picture is disappointing because the end of the spectrum
would correspond to the highest energy to which cosmic-ray acceleration
is possible, rather than to the Greisen-Zatsepin-Kuz'min effect in which
higher energy particles lose energy in interactions with
the cosmic microwave background~\cite{Greisen,ZK}.

\subsection{Comments on fitting with  several populations}

\begin{figure*}[thb]
\includegraphics[width=\columnwidth]{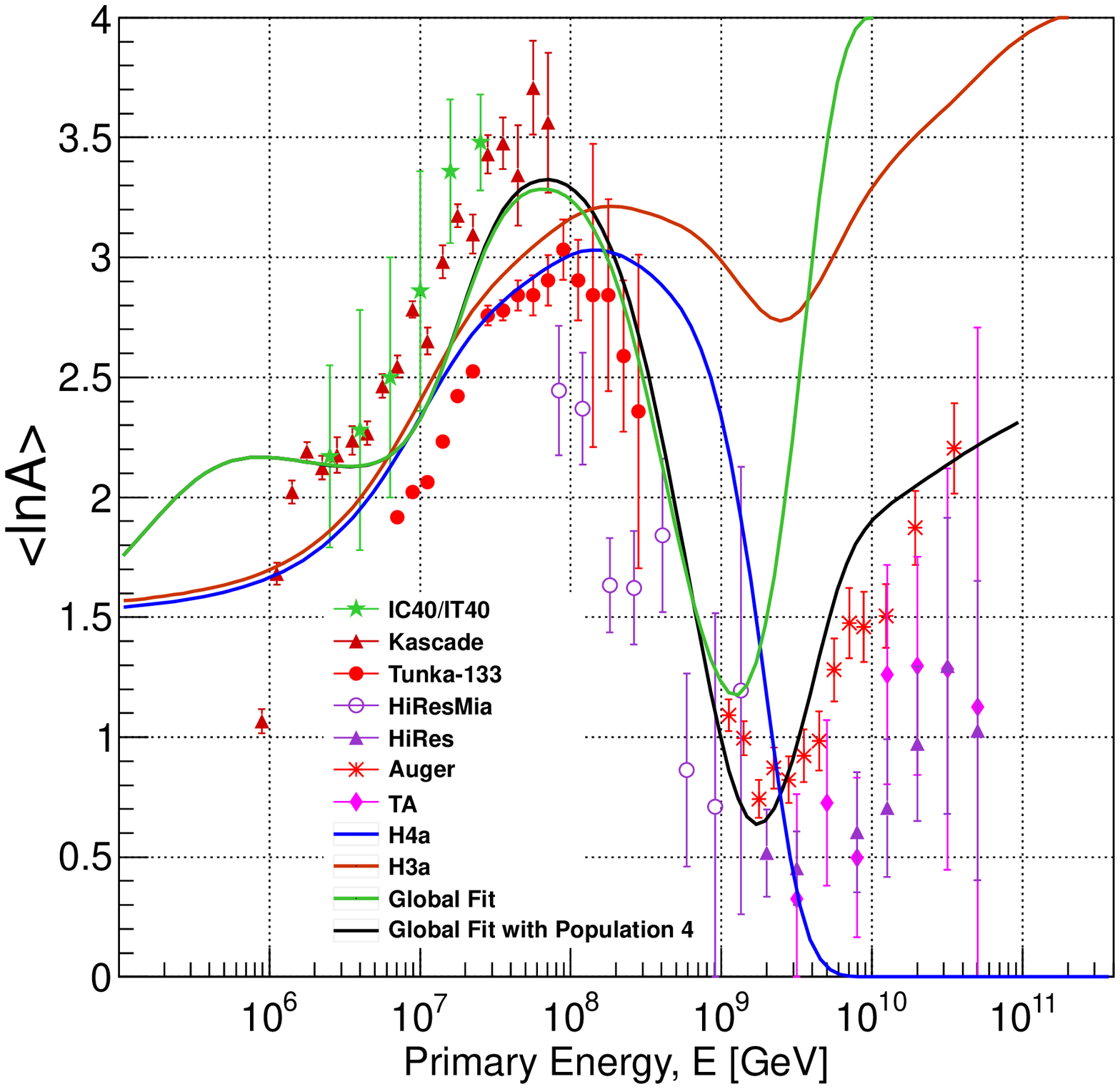}\includegraphics[width=\columnwidth]{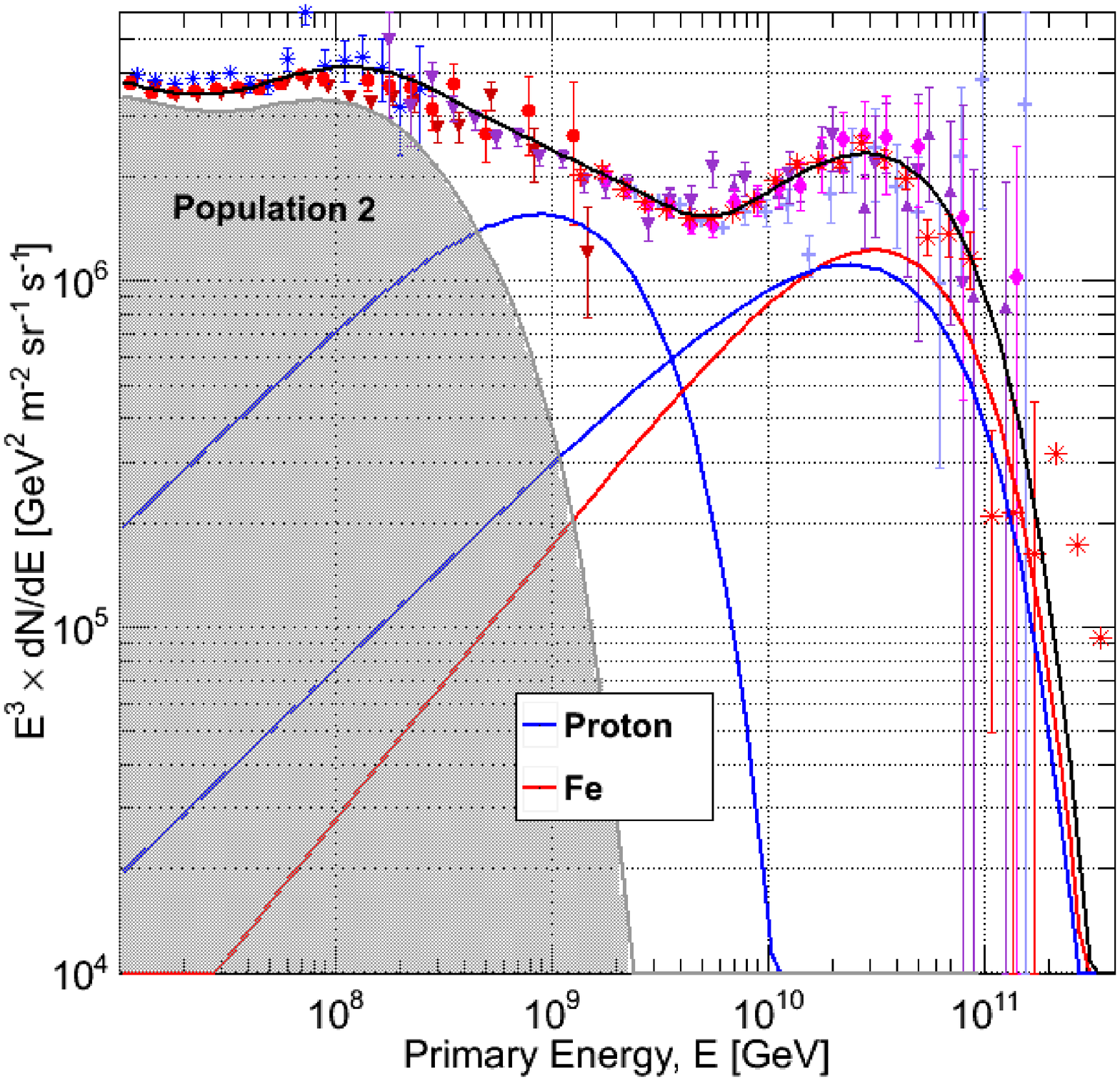}
  \caption{Left: Mean $\ln(A)$ for the four parameterizations
  of tables~\ref{tab2} and~\ref{tab3}.  (For iron $\ln(A)\approx 4.$)  Right: A modified fit with the addition
  of a 4th population of extra-galactic protons (see text for discussion).
  }
  \label{lnA}
 \end{figure*}

In both fits above we refer to three populations of particles, with spectral indices
for each nuclear component and a single characteristic maximum rigidity for each
population.  The latter assumption has the effect of making the composition become
heavier as each population approaches its maximum, as illustrated
in the left panel of Fig.~\ref{lnA}.  
Another important point is that the
higher energy populations can contribute significantly to the flux in the region
dominated by the lower population.  The right panel of Fig.~\ref{global} shows the overlap of
the three populations of the global fit of Table~\ref{tab3}.

The hardening of the spectrum observed by PAMELA and CREAM around $200$~GV
is suggestive of the onset of a new population~\cite{EW2}.  In this interpretation,
the Population 1 of our global fit would be a higher energy population
which becomes dominant above $200$~GV, but which still contributes 
significantly at lower energies.  Other explanations have been suggested.
For example, Ref.~\cite{PtuskinSeo} suggests that the hardening reflects
the concave spectrum characteristic of non-linear diffusive shock acceleration. 
In Ref.~\cite{Bing} it is suggested that a dispersion in the injection spectra
of different SNR is responsible for the hardening of the spectrum.  Reference~\cite{Blasi2}
shows how the hardening of the spectrum could be attributed to a change in
the type of turbulence responsible for diffusion of the cosmic rays.

A general feature illustrated by the various parameterizations discussed here
is that a Peters cycle of cutoffs of elemental components with rather hard
spectra before the cutoff can produce regions of the all-particle spectrum
that can be described approximately by steeper power laws.  The differential spectral index 
between 100~GeV and one PeV is close to 2.6 while the index above the
second knee, between $2\times10^{18}$ and $5\times10^{19}$~eV is approximately 3.35.
The individual spectra in the global fit of Table~\ref{tab3}, for example,
have differential indices below their cutoffs ranging from 2.2 to 2.4 (except for
hydrogen and helium below 200~GV).

In the case of the ankle structure, there is one model in which the absolute
energy of the feature is fixed by the physical assumptions
of the model.  That is the original work of
Berezinsky et al.~\cite{BGG}, which 
explains the dip in the plot of $E^3{\rm d}N/{\rm d}E$ as a consequence
of physical process of pair production by protons during propagation through the
cosmic microwave background radiation (CMB), which fixes the energy scale.  
In this ``dip'' model, the extragalactic
spectrum extends below the ankle and the galactic-extragalactic
transition occurs below one EeV.  In this case, according to 
Ref.~\cite{ABG} there is no need for
a second, higher energy galactic component B.  However,
in order to avoid a gap in the energy spectrum around $10^{17}$~eV,
the knee population would have to extend to significantly higher 
the $R_c\approx4$~GV as in both fits in this paper.

The different populations of particles presumably correspond to different
classes of sources.  For this reason it is instructive to compare
the energy content of the different populations of particles and
estimate the power required at the source.  As is well known,
the total energy in the cosmic-ray spectrum of galactic cosmic rays, which is dominated
by particles with energy below a TeV, can be provided by supernova explosions
at the rate of 3 per century.  The assumption is that approximately 10\%
of the kinetic energy released goes into acceleration of cosmic rays, presumably
by non-linear, first order, diffusive shock acceleration.  With $10^{51}$~erg
in kinetic energy of the ejecta per supernova explosion, the total power into cosmic rays is
then $\sim 3\times10^{50}$~erg/century or $10^{41}$~erg/s.

It is interesting to compare the power requirement for the second galactic population in
the two models described above with the total power of the galactic cosmic-ray sources.  To estimate
this from the parameterizations of Tables~\ref{tab2} and \ref{tab3}, we start from
a simplified version of the diffusion equation,
\begin{equation}
N(E) \;=\;Q(E)\times \tau_{\rm esc}(E).
\label{diffusion}
\end{equation}
Here $N(E)$ is the density of cosmic-ray particles (differential in energy)
and $Q(E)$ is the number of particles per second per unit volume at which the sources
inject particles of energy $E$.  $\tau_{\rm esc}(E)\,=\,\tau_0 E^{-\delta}$ is the
energy dependent escape time from the galaxy.  We assume $\tau_0\,=\,10^7$~yrs
and $\delta\,=\,0.33$.~\footnote{The measured ratio of secondary/primary
nuclei at low energy decreases like $E^{-0.6}$~\cite{Obermeier}.  Such a strong energy
dependence cannot continue to the knee region without producing unobserved anisotropy in
the cosmic radiation at high energy.}  Multiplying by the factor $4\pi/c$, which
converts flux to density, we can then integrate the spectra for Population 2
given the parameters in the Tables.  We find $6\times 10^{48}$~erg/century
for Component B (Population 2) of the Hillas model as parameterized in Table~\ref{tab2} and 
$2\times 10^{49}$~erg/century for Population 2 of the global fit of Table~\ref{tab3}.
Both are reasonable in the sense that they could be accounted for by special sources
at a level less than 10\% of the total power requirement for all galactic cosmic rays.

\subsection{Composition}

Determination of composition with indirect air shower experiments is difficult,
primarily because of the large fluctuations from shower to shower, which tend
to smear out differences arising from the different mixture of primaries.
Three data samples are shown in the left panel of Fig.~\ref{lnA} below $10^{18}$~eV.
TUNKA results~{\cite{TUNKA} are from measurement of the lateral distribution of Cherenkov light
on the ground.  IceCube results are from an analysis of one month of
data on coincident events obtained when the detector was half complete~\cite{IceCube}.
Coincident events are those in which nearly vertical showers are reconstructed both
by IceTop on the surface and by the deep detectors of IceCube more than 1.5 km
below the surface.
The composition-sensitive parameter is the ratio of the ~$\sim$TeV muons in the shower
core to the shower size at the surface.  Heavier primaries produce more light
due to muon energy loss in the deep detector for a given shower size at the surface
compared to light primaries.  The KASCADE result~\cite{Kascade} is based on
measurements of the ratio of electrons to muons in showers at the surface, and
we take the $<\ln(A)>$ values as plotted in Ref.~\cite{KU}.  

In their review~\cite{KU} Kampert \& Unger interpolated the depth of
maximum measurements of several air fluorescence detectors
between predictions for a pure proton assumption and a pure iron assumption
in order to obtain a value for $<\ln(A)>$.  We show in the left panel of Fig.~\ref{lnA}
the values they inferred from HiRes-MIA~\cite{HiRes-MIA}, Auger and Telescope array.
  The results depend to some extent on the hadronic
interaction model used to calculate depth of maximum for protons and iron,
but the trend of the data is similar in different models.  We plot their
results for the SIBYLL model~\cite{SIBYLL}.

An important early result was obtained by the prototype of the HiRes fluorescence
telescope observing showers in coincidence with the underground muon array
that formed the muon detector for CASA-MIA.  The combined hybrid detector allowed
the profile of nearby, relatively small showers to be reconstructed.  The
measured depth of maximum distribution they observed~\cite{HiRes-MIA}
showed the depth of maximum increasing rapidly from $10^{17}$~eV
to $3\times 10^{18}$~eV in a manner consistent with all the parameterizations
in the left panel of Fig.~\ref{lnA} except the disappointing model with
only iron at the highest energies.

There is a significant disagreement in interpretation of depth of maximum
measurements above $10^{18}$~eV, with Hi-Res and TA preferring nearly pure
protons and Auger preferring a transition to heavies.  But in both cases,
a transition to pure iron, as would be the case at the end of the spectrum
in the case of the disappointing model, does not seem to be indicated.
The right panel of Fig.~\ref{lnA} shows a fit that contains a 4th population
made entirely of protons to show the amount that would be needed
to bring the composition closer to the data in this energy region.
There would need to comparable amounts of protons and iron.  
We note that such a mixture would not be consistent with the small fluctuations
seen by Auger~\cite{AugerFluctuations}, which would seem to require a more pure 
composition of heavy nuclei.  It appears that there is not at present
a satisfactory understanding of the highest energy cosmic rays.

\section{Conclusion}

\subsection{Transition from galactic to extragalactic sources}

  The energy range were Populations 2 and 3 meet is usually treated as the
 transition from galactic to extragalactic cosmic ray sources. There are
at least three different ways in which this transition region can be treated. 

  In the dip model~\cite{BGG} the transition region is below 10$^{18}$ eV.
 The dip itself is generated by the energy loss on production of
 electron-positron pairs in interactions on the microwave background.
 The energy spectrum of the extragalactic protons is flatter below 
 10$^{18}$ eV. What that means is the transition from heavy galactic
 nuclei to extragalactic protons and He nuclei happens below that energy.
 This is very different from the classical `ankle'
 model~\cite{Waxman95}  where
 the transition region is above that energy and the galactic cosmic ray
 contribution above 10$^{19}$ eV is significant. The chemical composition 
 of cosmic rays then starts changing significantly only above 10$^{19}$ eV. 

 The third model is that of mixed cosmic ray composition at the highest
 energies~\cite{Allard2012}.
 In such models the extragalactic cosmic rays are accelerated with a 
 chemical composition similar to that of the GeV galactic cosmic rays.
 Their composition is changed in propagation because of photoproduction,
 electron positron pair creation, and nuclear photodisintegration.
 The transition between galactic and extragalactic cosmic
 rays is at intermediate energy, and the final energy spectrum and 
 chemical composition depend strongly on the distance to the
 extragalactic sources.

\subsection{Outlook}

As noted in the previous paragraphs, there is much interest in
studying the transition from galactic to
extragalactic cosmic rays, which is expected in the energy range
between 10$^{17}$ and 10$^{19}$~eV.  There is a lack of data
at present because this energy range is at the upper
end of the range of kilometer-scale experiments such as
TUNKA, KASCADE-Grande and IceTop, and it overlaps the threshold
region for the giant air shower detectors, Auger and Telescope Array.
As a consequence, we do not yet have a good understanding of
this important transition region.  Only preliminary results from
IceCube and IceTop have been presented so far.  With the completion
of the detector and operation with the full detector since
May, 2011, we can expect to see significant new results from IceCube
at the beginning of the transition region, but the kilometer-scale
detectors will always be statistically limited above the EeV energy range.
From the higher energy side, the big experiments
 are working to decrease their energy threshold and become 
 efficient down to the 10$^{17}$~eV energy range. 
 
 The Auger Southern Observatory
 has added
 the High Elevation Auger Telescope (HEAT, \cite{HEAT}) and
 the Auger Muon and Infill for the Ground Array (AMIGA, \cite{AMIGA})
 to the original design in order to lower their threshold.
Air showers of energy around 10$^{17}$~eV  
 do not emit sufficient light to be seen further
 than a few kilometers away.  At that distance the depth of
 maximum $X_{max}$ appears  at higher elevation, 
 outside of the viewing angle of the original design.
 In addition, nearby showers of lower energy reach their maximum of development
 above the  30$^\circ$ limit of the original Auger telescopes.
 HEAT is composed of three fluorescence
 telescopes of the same basic design as the original Auger telescope
 and is  installed  at the western fluorescence detector site
 of the observatory. They can operate in two positions.
 Horizontally they share the same field of view as the original telescopes.
 This position is used for calibration of the instruments.
 Tilted upward by 29$^\circ$, which is the normal operation mode
 for HEAT, they observe the upper part of the atmosphere.
 The first light from one of those telescopes was seen in
 January 2009.

 An infill array of 85 detectors is deployed
 on two grids of one half (750~m) and one fourth (433~m) of
 the regular Auger surface array grid over the field of view of
 HEAT. The 750 m infill covers an area of 23.5 km$^2$ and the
 433 m infill covers 5.9 km$^2$. There are also plans to have 
 muon counters in AMIGA.  The close spacing of the
 new subarray is $10^{17}$~eV.
 
 The Telescope Array experiment is in the process of building
 its low energy extension TALE (G.~Thomson, {\em private communication}).
 It will consist of ten fluorescent telescopes located at the Northern
 side of the array. These telescopes will be elevated to be sensitive 
 from 31$^o$ to 59$^o$ above the horizon.
 One hundred new scintillator counters will be deployed 
nearby at distances
 of 400 m from each other (1/3 of the standard spacing in the Telescope
 Array).  Thirty-five of them will be operational in March 2013.
 TALE will be sensitive to primary particles of
 energy above 10$^{16.5}$~eV.
 
 In summary, a thorough study with good statistics of the energy region
 between $10^{17}$ and $10^{19}$~eV should be scientifically productive.

\end{document}